\documentclass{elsart}
\usepackage{amssymb,amsbsy}
\usepackage{graphicx}   

\def\be{\begin{equation}}
\def\ee{\end{equation}}
\def\bea{\begin{eqnarray}}
\def\eea{\end{eqnarray}}
\def\bra#1{\left\langle #1\right|}
\def\ket#1{\left| #1\right\rangle}
\def\bpi{\boldsymbol\pi}
\def\btau{\boldsymbol\tau}

\def\bphi{\boldsymbol\phi}

\def\Jl#1#2#3#4{{#1} {#2} (19#3) #4}
\def\NPA{Nucl. Phys. A}
\def\NPB{Nucl. Phys. B}
\def\PLB{Phys. Lett. B}
\def\PRL{Phys. Rev. Lett.} 
\def\PREP{Phys. Rep.} 
\def\PRC{Phys. Rev. C}
\def\PRD{Phys. Rev. D}

\begin{document}

\begin{frontmatter}

\title{The role of the Roper resonance in
$\mathbf{n p \rightarrow d \boldsymbol{(}\bpi \bpi\boldsymbol{)}^0}$}

\author{L. Alvarez-Ruso}
\address{Departamento de F\'{\i}sica Te\'{o}rica and IFIC, Centro Mixto 
Universidad de Valencia-CSIC, 46100 Burjassot, Valencia, Spain}

\begin{abstract}
In this work, a model for the $n p \rightarrow d (\pi \pi)^0$ reaction is
developed. It is shown that the structure of the deuteron momentum spectra
for a neutron beam momentum of 1.46 GeV 
can be explained as a consequence of the interplay of two mechanisms involving
the excitation of the $N^*(1440)$ resonance and its subsequent decay into
$N (\pi \pi)^{T=0}_{S-wave}$ and $\Delta \pi$ respectively. 
The relevance of the present
analysis for the study of the Roper excitation and decay properties, as well
as for the interpretation of other two-pion production experiments is
discussed. 
\end{abstract}

\end{frontmatter}

\section{Introduction}

The study of nucleon-nucleon inelastic collisions provides a powerful tool to
deepen our insight into the properties of nucleon-nucleon interactions and
baryonic resonances. A large amount of theoretical work on threshold meson
production has been performed over the last years \cite{teo}, 
stimulated by the precise data obtained at IUCF, CELSIUS and COSY \cite{exp}. 
On the other side, double pion production reactions in the nucleon,
$\gamma N \rightarrow N \pi \pi$ and $\pi N \rightarrow N \pi \pi$,
have proved to be essential as a test of Chiral Perturbation Theory  
\cite{meissrep} and as a source of information about $N^*(1440)$ and
$N^*(1520)$ \cite{jantonio,manley,manolo,jensen}. In this context, the still 
scantily explored two pion production channel in the collisions of nucleons 
and light nuclei appears as promising research area. 

Most of the work, both experimental and theoretical, on two pion production in
nucleon-nucleon collisions was performed in the seventies and in connection
with the ABC effect. The ABC anomaly is an enhancement in the missing mass
spectra close to the $\pi \pi$ production threshold, observed for the reactions
$p d \rightarrow {^3}He X$ \cite{abc}, $n p \rightarrow d X$ \cite{plouin} and
$d d \rightarrow {^4}He X$ \cite{banaigs}. Although any interpretation of the
ABC as a resonance is excluded \cite{plouin}, the origin of it is still poorly
understood \cite{plouin,wilkin}. An important step towards the understanding
of the ABC effect has been taken in Ref. \cite{anders}, where the ${^4}He$
spectra from the $d d \rightarrow {^4}He X$ reaction at a deuteron beam energy 
of 1250 MeV \cite{banaigs} has been explained assuming that pions are
independently produced in reactions involving two different pairs of nucleons 
from the projectile and target deuterons. Nowadays, two pion 
production in $p p$ collisions is  being studied experimentally at CELSIUS 
for beam kinetic energies between 650
and 775 MeV. On the theoretical side, a microscopic model for the
$N N \rightarrow N N \pi \pi$ reaction has been recently developed \cite{yo}. It
includes mechanisms with the excitation of $N^*$ and $\Delta$ resonances, as
well as some non-resonant contributions, and gives a satisfactory description
of the available experimental data on total cross sections for most of the 
channels and in a wide range of energies up to 800 MeV above threshold.

In this letter, I focus the attention on the deuteron
spectrum in $n p \rightarrow d (\pi \pi)^0$ measured by Hollas and
collaborators \cite{hollas} using a nearly monokinetic neutron beam with central
momentum $p_n = 1.463$ GeV. This experiment is somewhat similar to the one of
Plouin et al. \cite{plouin}, but at lower energies ( $T_n = 795$ MeV in 
\cite{hollas} vs $1160$ MeV in \cite{plouin} ). Therefore, the analysis is
simpler since, once the $\pi^0$ peaks are subtracted, only the double pion
production mechanism is present. Apart from that, one expects that, being
closer to threshold, the reaction mechanism might be simpler. The ABC peaks
are not present in the data; they rather show a well defined bump at high
$\pi \pi$ missing masses, in disagreement with the models available in the
literature \cite{levy,bnrs}. From this comparison, the authors concluded 
\cite{hollas} that neither double-$\Delta$ formation nor double-nucleon
exchange provides the appropriate description of the reaction at 
$p_n = 1.46$ GeV.  A similar enhancement has also been observed for 
the reaction $p d \rightarrow {^3}He \pi^+ \pi^-$, which is being studied 
using a beam of protons from COSY ( MOMO experiment ) \cite{momo}, at a Q 
value close to the one of the experiment by Hollas et al. \cite{hollas}. 
These common features point to a common dynamical description.

Here, it is shown that the deuteron spectra for
$n p \rightarrow d (\pi \pi)^0$ at $p_n = 1.46$ GeV can be understood as a
consequence of the interference of two mechanisms involving the excitation of
the Roper resonance $N^*(1440)$ and its subsequent decay into
$N (\pi \pi)^{T=0}_{S-wave}$ and $\Delta \pi$ respectively. The implications
of this finding to the MOMO experiment are also discussed.

\section{The model}

\begin{figure}[h!]
\begin{center}
\includegraphics[height=\textwidth, angle=-90]{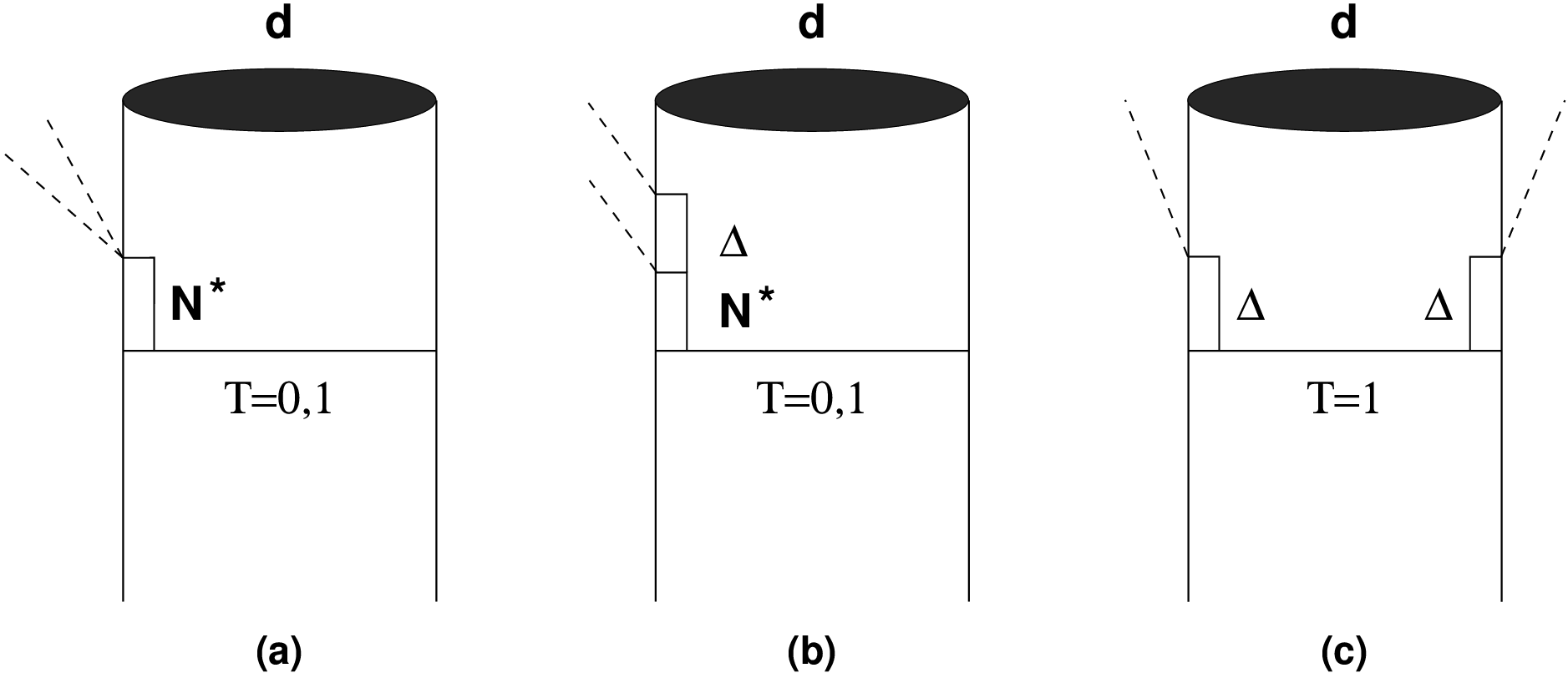}
\caption{Set of diagrams of the model.}
\label{diag}
\end{center}
\end{figure}

The model is schematically  presented in Fig. \ref{diag}. It is a reduced
version of the model of Ref. \cite{yo}, modified for the case where one has a
deuteron instead of two free nucleons in the final state. The choice of the 
mechanisms was based on their contribution to the total cross section for the
$p n \rightarrow p n \pi^+ \pi^-$ reaction; the situation for the 
$p n \rightarrow p n \pi^0 \pi^0$ channel is similar. At $T_p = 800$ MeV, the 
mechanism with $N^* \rightarrow N (\pi \pi)^{T=0}_{S-wave}$ gives
$\sigma \sim 11\, \mu b$, being by far the most important. The second largest
contribution comes from  $N^* \rightarrow \Delta \pi$ with
$\sigma \sim 0.5\,\mu b$; as we will see, in the case of a deuteron in the
final state, its contribution is larger with respect to the dominant
$N^* \rightarrow N (\pi \pi)^{T=0}_{S-wave}$ and crucial to obtain the
right shape. All other mechanisms give $\sigma \lesssim 0.3\, \mu b$; I do
not include them all, but just the double-$\Delta$ ( $\sigma \sim 0.1\, \mu b$ ) 
one in order to make contact with the model of Ref. \cite{bnrs}.  

The deuteron momentum spectrum is the sum of the $\pi^+ \pi^-$ and the 
$\pi^0 \pi^0$ contributions. In the Laboratory frame, the charged pions piece
is given by 

\be
\label{eq1}
\frac{d {^2}\sigma}{d p'_d d \Omega'_d} = \frac{1}{4} \frac{1}{(2 \pi)^5}
\frac{M M_d (p'_d)^2}{E'_d p_n p_d} \left( \int d E_\pi d \varphi_\pi 
\frac{1}{4} \sum_{R r_1 r_2} |\mathcal{M}_{R r_1 r_2}|^2 \right)_{CM} \,. 
\ee
Here, $E'_d$ and $p'_d$ are the deuteron energy and the modulus of its
momentum, both in Lab. frame; $p_d$ is the modulus of the deuteron momentum
in the center-of-mass system (CM); $M$ and $M_d$ stand for the nucleon and
deuteron masses respectively. The integral in brackets must be calculated
with all the kinematical variables defined in CM; it runs over the polar
angle and the energy of one of the outgoing pions. The amplitude squared
is summed over the deuteron spin ($R$) and averaged over the spins of the
incoming nucleons ($r_1$, $r_2$). For the neutral pions channel, the
expression is the same but with and extra $1/2$ factor, which is a
consequence of having two identical pions in the final state. The difference
of masses between charged and neutral pions is taken into account for the
phase space, but isospin symmetry is assumed in the calculation of the
amplitude. 

The amplitude can be expressed as 

\bea
\label{eq2}
&\mathcal{M}_{R r_1 r_2}= \sum_{r'_1 r'_2} \left(\frac{1}{2} r'_1 
\frac{1}{2} r'_2 \right| \left. \vphantom{\frac{1}{2}} 1 R\right) \int 
\frac{d \mathbf{q}}{(2 \pi)^3} D_{T=0,1}(q) \times \nonumber\\[.4cm]
&\left\{ 
\left( \bra{p r'_1} \hat{V}_1 \ket{p r_1} \bra{n r'_2} \hat{V}_2 \ket{n r_2} 
- \bra{n r'_1} \hat{V}_1 \ket{p r_1} \bra{p r'_2} \hat{V}_2 \ket{n r_2}
\right) \tilde{\varphi}_d(\mathbf{P_2}) \right. \nonumber\\[.4cm]
&+ \left. 
\left( \bra{p r'_1} \hat{V}_2 \ket{p r_1} \bra{n r'_2} \hat{V}_1 \ket{n r_2} 
- \bra{n r'_1} \hat{V}_2 \ket{p r_1} \bra{p r'_2} \hat{V}_1 \ket{n r_2}
\right) \tilde{\varphi}_d(\mathbf{P_1}) \right\} 
\eea
where $\tilde{\varphi}_d(\mathbf{P})$ is the s-wave deuteron wave function in
momentum space, normalized as 

\be
\label{eq3}
\int \frac{d \mathbf{k}}{(2 \pi)^3}\, {\tilde{\varphi}_d}^2(\mathbf{k}) = 1.
\ee
The d-wave part has been neglected. For the wave function, different
expressions and parameterizations can be used \cite{paris,bonn,hulthen}. 
The value of $\mathbf{P_{1(2)}}$ depends
on the mechanism; for those of Figs. \ref{diag}a and \ref{diag}b,
$\mathbf{P_{1(2)}} = \mathbf{q}+\mathbf{p_{1(2)}}-\mathbf{p_d}/2$ and for the
$\Delta-\Delta$ mechanism ( Fig. \ref{diag}c )
$\mathbf{P_{1(2)}} =
\mathbf{q}+\mathbf{p_{1(2)}}-\mathbf{p_d}/2-\mathbf{p}_{\bpi}$,
$\mathbf{p_{1(2)}}$ and $\mathbf{p_d}$ been the momenta of the proton
(neutron) and deuteron respectively; $\mathbf{p}_{\bpi}$ is the momentum of
the pion, over whose energy the integral in Eq. \ref{eq1} is performed.
The function $D_{T=0,1}(q)$, which stands for the meson propagators and form
factors, will be discussed later; $q = (q_0, \mathbf{q})$  is the four
momentum transfer from one nucleon to the other; $q_0$ is given by energy
conservation in the vertices and, therefore, depends on the energy of one
of the outgoing nucleons, taken to be one half of the deuteron energy. 

The matrix elements in Eq. \ref{eq2} are evaluated for the different
mechanisms using the Feynman rules that can be obtained using
phenomenological Lagrangians; some of the required ones are

\be
\label{eq4}
{\mathcal L}_{\Delta N \pi} = \frac{f^*}{m_\pi} \psi^{\dagger}_\Delta 
S^{\dagger}_i (\partial_i\bphi) \mathbf{T}^{\dagger} \psi_N \, + \, h. c. \,,
\ee
\be
\label{eq5}
{\mathcal L}_{N^* N \pi}
= \frac{\tilde{f}}{m_\pi} \psi^{\dagger}_{N^*} \sigma_i (\partial_i
\bphi) \btau \psi_N  \, + \, h. c. \,,
\ee
\be
\label{eq6}
{\mathcal L}_{N^* \Delta \pi} = \frac{g_{N^* \Delta \pi}}{m_\pi}
\psi^{\dagger}_{\Delta} S^{\dagger}_i (\partial_i \bphi) \mathbf{T}^{\dagger} 
\psi_{N^*} \, + \, h. c. \,.
\ee
In Eqs. \ref{eq4}, \ref{eq6}, $\mathbf{S^{\dagger}}$ ($\mathbf{T^{\dagger}}$) 
are the spin (isospin) $1/2 \rightarrow 3/2$ transition operators  
\cite{erweise}; $\psi_N$, $\psi_{\Delta}$, $\psi_{N^*}$ and $\bphi$ stand for
the nucleon, Delta, Roper and pion fields, while $m_\pi$ is the pion
mass. The absolute value of the coupling constants
$f^* = 2.13$, $\tilde{f} = 0.477$ and $g_{N^* \Delta \pi} = 2.07$ are obtained
from the partial decay widths of the $\Delta$ and $N^*(1440)$ \cite{pdg}. In
the case of the decays $N^* \rightarrow N \pi$ and
$N^* \rightarrow \Delta \pi$, branching ratios of 65$\%$ and 25$\%$
respectively are assumed, as well as an $N^*$ total width of 350 MeV; the
signs correspond to those provided in earlier analyses of the ($\pi, \pi \pi$) 
reactions \cite{manley,manolo}. The Lagrangian for the $N N \pi$ vertex is the
standard one which, in the non-relativistic limit, looks like

\be
\label{eq6a}
{\mathcal L}_{N N \pi}
= \frac{f_{N N \pi}}{m_\pi} \psi^{\dagger}_N \sigma_i (\partial_i
\bphi) \btau \psi_N \,,
\ee
with $f_{N N \pi} = 1$.

A general Lagrangian for the $N^* \rightarrow N (\pi \pi)^{T=0}_{S-wave}$
decay \cite{meiss2pi} is 

\bea
\label{eq7}
{\mathcal L}_{N^* N \pi \pi} = 
&-& c^*_1 \frac{m^2_\pi}{f^2} \bar{\psi}_{N^*} {\bphi}^2 \psi_N 
\nonumber\\[.2cm] 
&-& c^*_2 \frac{1}{f^2 {M^*}^2} (\partial_\mu \partial_\nu 
\bar{\psi}_{N^*})(\btau \partial_\mu \bphi) 
(\btau \partial_\nu \bphi)\psi_N \, + \, h. c.
\eea
where $f = 92.4$ MeV is the pion decay constant and $M^*$, the mass of the
Roper resonance. Using the partial decay width, the parameters $c^*_1$ and
$c^*_2$ can be constrained to an ellipse \cite{meiss2pi}. In order to further
constrain them, the model of Ref. \cite{manolo} and the overall data on
$\pi^- p \rightarrow \pi^+ \pi^- n$ have been used \cite{yoproc}. Assuming a
branching ratio of 7.5$\%$ for the $N^* \rightarrow N (\pi \pi)^{T=0}_{S-wave}$ 
decay, the best agreement is obtained for
$c^*_1 = -7.3$ GeV$^{-1}$ and $c^*_2 = 0$ (Set I), but the data seem to be
still compatible with the choice of $c^*_1 = -12.7$ GeV$^{-1}$ and
$c^*_2 = 2.0$ GeV$^{-1}$ (Set II). In this study, Set I will be used except
where a different choice is explicitly indicated. 

The matrix elements of Eq. \ref{eq2} contain, apart from the vertices 
described above, Roper and Delta propagators, given by

\be
\label{eq8}
D_l(p)=\frac{1}{\sqrt{p^2}- M_l+ \frac{1}{2} i \Gamma_l(p)} 
\frac{M_l}{\sqrt{M_l^2+\mathbf{p}_l^2}} \quad ; \qquad l=(N^*,\Delta)
\ee
with $p = (p_0, \mathbf{p})$ the momentum of the resonance and
$\Gamma_l(p)$, its total width. The partial decay $\Delta \rightarrow N \pi$
practically accounts for the total Delta width. In the case of the Roper, 
the major part
of the width comes from the decay into nucleon and pion and the rest from
the decay into nucleon and two pions. All of them, except the small (less
than 8$\%$ at the $N^*$ peak) $N^* \rightarrow N \rho$ decay, can be 
calculated with the Lagrangians given above.  

Finally, let us consider $D_{T=0,1}(q)$. For the $T=1$ potential, I
calculate the diagrams assuming a pion exchange and make the substitution 

\be
\label{eq9}
D^{(\pi)}_{T=1} (q) q_i q_j \rightarrow 
V'_{L} (q) \hat{q}_i \hat{q}_j +  V'_{T} (q) 
( \delta_{ij} - \hat{q}_i \hat{q}_j )
\ee
so that $D_{T=1}$
includes the longitudinal pion exchange, the transversal rho exchange, and
the short range correlations that take into account the repulsive force at
short distances. Functions $V'_{L} (q)$ and $V'_{T} (q)$ are described
elsewhere \cite{yo,weise}. With respect to the $T = 0$ channel, in a recent
analysis \cite{zaki} of the $(\alpha, \alpha')$ reaction on a proton target, the
strength of the isoscalar $N N \rightarrow N N^*$ transition was extracted by
parameterizing the transition amplitude in terms of an effective ``$\sigma$''
which couples to $N N$ as the Bonn model $\sigma$
($g^2_{\sigma N N}/ 4 \pi = 5.69$) \cite{bonn} and couples to $N N^*$ with an 
unknown strength provided by a best fit to the data
($g^2_{\sigma N N^*}/ 4 \pi = 1.33$). According to the fit, $g_{\sigma N N}$
and $g_{\sigma N N^*}$ have the same sign. Therefore, we have

\be
\label{eq10}
D_{T=0} (q) = \frac{1}{q^2 - m^2_{\sigma}} 
\left(\frac{\Lambda^2_\sigma - m^2_{\sigma}}{\Lambda^2_\sigma - q^2} \right)^2
\,,
\ee
where an equal $\Lambda_\sigma = 1.7$ GeV for the form factor is assumed for
both vertices. The explicit expressions for the amplitudes are too involved
to be reproduced here, but some details of their evaluation can be traced in
Ref. \cite{yo}, where all the amplitudes for the free reaction
$p p \rightarrow p p \pi^+ \pi^-$ are given in the Appendix.

\section{Results and discussion}

\begin{figure}[h!]
\begin{center}
\includegraphics[height=\textwidth, angle=-90]{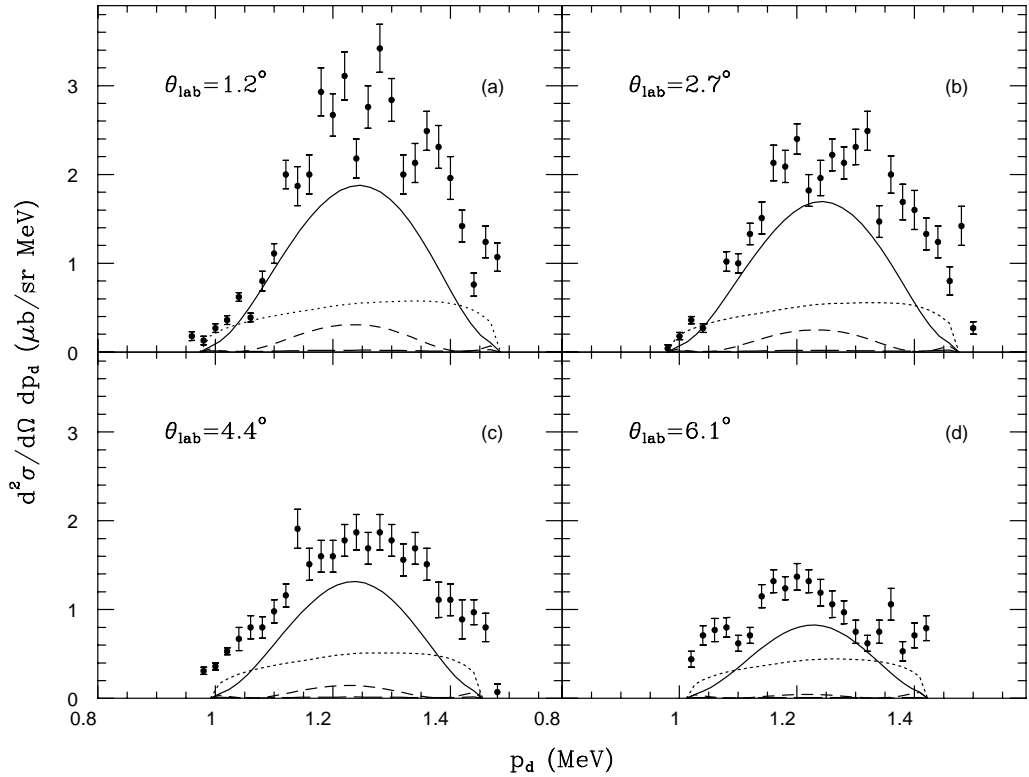}
\caption{Deuteron momentum spectra for $n p \rightarrow d (\pi \pi)^0$
at $p_n = 1.46$ GeV and different laboratory angles (solid lines) compared
to the measured data \cite{hollas}. The dotted line
corresponds to the $N^* \rightarrow N (\pi \pi)^{T=0}_{S-wave}$ mechanism
(Fig. \ref{diag} a); the short-dashed line stands for the
$N^* \rightarrow \Delta \pi$  (Fig. \ref{diag} b) and the long-dashed one, for
the double-$\Delta$ (Fig. \ref{diag} c).}
\label{all}
\end{center}
\end{figure}

With the ingredients described in the previous section, one can calculate
the deuteron spectra at $p_n = 1.46$ GeV for different angles.
The results are shown in Fig. \ref{all}. They
compare quite well with the data of Hollas et al. \cite{hollas}, and certainly
much better than the previous models \cite{levy,bnrs}. The curves, in
general, underestimate the data, maybe because there are many other
mechanisms, not considered for the sake of simplicity, that, even been
individually small, could in sum enhance the cross section. Some of the
approximations made, like the neglect of the deuteron d-wave and
non-relativistic approximation in the vertices, as well as the intrinsic
uncertainties of the mechanisms included can also be a source of discrepancies;
they are discussed below. The
large data point at the edges of the spectra show the contamination of the
$\pi^0$ peaks \cite{hollas}. I have also estimated the influence of the width
of the neutron beam by averaging the double differential cross sections over
a Breit-Wigner profile of 40 MeV width and centered at $p_n = 1.46$ GeV
\cite{hollas}; the contribution of this effect is very small and irrelevant 
for our study. The plotted curves are 
obtained using the deuteron wave function derived from the Paris
potential \cite{paris}; with the Bonn wave functions \cite{bonn} the results
are very similar, while for the Hulthen one \cite{hulthen}
the distributions are overall larger, up to a factor two at the position of 
the central bump. 

The mechanism $N^* \rightarrow N (\pi \pi)^{T=0}_{S-wave}$ (Fig. \ref{diag}a)
produces spectra very similar to phase space; its contribution is certainly the
largest, but its relative size with respect to the $N^* \rightarrow \Delta \pi$
mechanism is not as large as one would naively expect from estimations based on
the total cross sections obtained for both mechanisms in the free
$N N \rightarrow N N \pi \pi$ reaction. Nevertheless, it is not surprising
since here we are sensitive only to a reduced phase space region and choosing
a particular combination of the quantum numbers of the outgoing nucleons
(those of the deuteron). For this mechanism, I have calculated the
contribution of the d-wave part of the deuteron wave function and found it
negligible. 

\begin{figure}[h!]
\begin{center}
\includegraphics[height=0.5\textwidth,angle=-90]{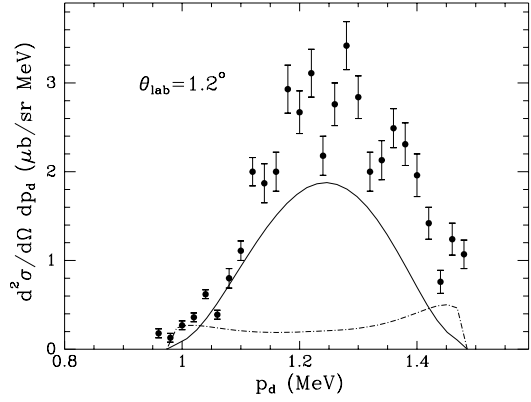}
\caption{Calculated spectra for two different
choices of the $g_{N^* \Delta \pi}$ sign. The data clearly favor the positive
sign (solid line) with respect to the negative (dash-dotted line).}
\label{signo}
\end{center}
\end{figure}

The $N^* \rightarrow \Delta \pi$ mechanism (Fig. \ref{diag}b) exhibits a wide
bump at the center of the spectra (high $\pi \pi$ masses), whose maximum falls
fast with the increase of the deuteron angle, and small peaks at the edges
of the spectra (low $\pi \pi$ masses); the size of these peaks does not vary
appreciably with the angle. This mechanism plays a crucial role in providing
the right shape to the distributions through its interference with the larger
$N N \rightarrow N N \pi \pi$ contribution. This interference is constructive
at high $\pi \pi$ masses and destructive at low ones. Such pattern can be
understood by realizing that the $N^* \rightarrow \Delta \pi$ amplitude is
dominated by terms proportional to the scalar product of the outgoing pions
three momenta; this scalar product has different signs in the center of the
spectra, where the pions go back to back, and at the edges, where they
travel together. In order to further illustrate the effect of the interference,
Fig. \ref{signo} shows the effect of changing the relative sign of the two
amplitudes. The data clearly favor a choice of the sign of 
$g_{N^* \Delta \pi}$ in agreement with earlier works \cite{manley,manolo}.

\begin{figure}[h!]
\begin{center}
\includegraphics[height=0.5\textwidth,angle=-90]{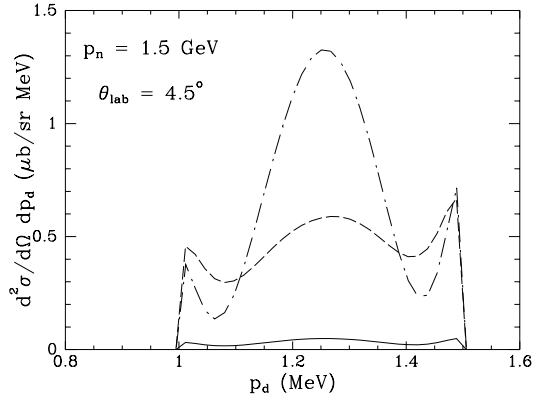}
\caption{Contribution of the double-$\Delta$ mechanism in the case of $\pi$
exchange alone (dashed line), $\pi + \rho$ exchange (dash-dotted line) and
$\pi + \rho +$ short range correlations (solid line). In this case the
calculation uses a Hulthen wave function for the deuteron.}
\label{deldel}
\end{center}
\end{figure}

The double-$\Delta$ mechanism (Fig. \ref{diag}) is so small that it can hardly
be distinguished in Fig. \ref{all}. In Fig. \ref{deldel}, the contribution of
this mechanism alone is shown for $p_n = 1.5$ GeV, $\theta_{lab}=4.5^{\circ}$
and using the Hulthen wave function, in order to compare with the result of
Bar-Nir, Risser and Shuster (Fig. 4 b of Ref. \cite{bnrs}). The differential
cross section obtained in the case of only pion exchange  is very similar
to the one given by the relativistic model of  Ref. \cite{bnrs}; the
inclusion of the rho exchange modifies the result, and the short range
correlations between the initial nucleons
cause a strong reduction of the strength of this mechanism.
At $T_p = 775$ MeV ($p_p = 1.43$ GeV), the $p p \rightarrow p n \pi \pi$
and $p p \rightarrow d \pi \pi$ reactions are probably dominated by the 
double-$\Delta$ mechanism; therefore, the future data from CELSIUS would
provide an excellent opportunity to study the interplay of the described
ingredients and, in particular, the role of the short range correlations.

\begin{figure}[h!]
\begin{center}
\includegraphics[height=\textwidth,angle=-90]{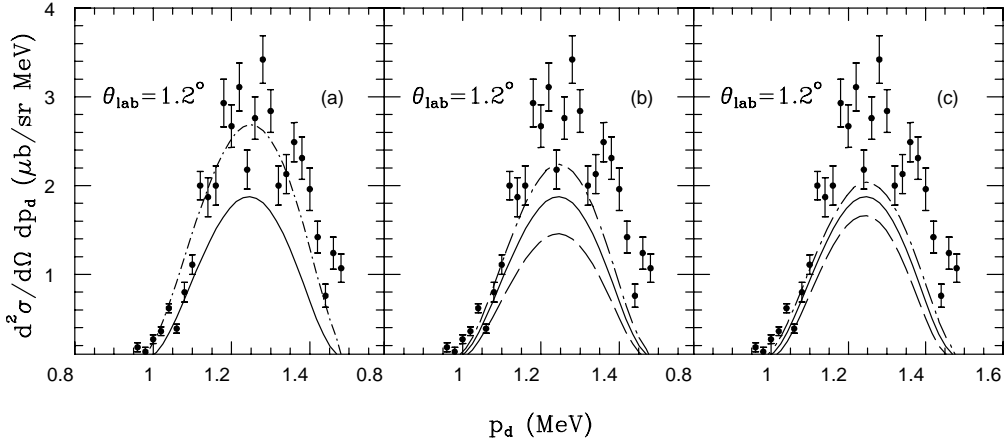}
\caption{Dependence of the spectrum on certain decay properties of the
$N^*(1440)$. The solid line in (a) shows the result for the Set I of
parameters $c^*_1$, $c^*_2$ used everywhere else ($c^*_1 = 0$), and the
dash-dotted one is obtained for Set II. In Fig. (b), different values of the 
$N^*$ total width are considered: dashed line, 250 MeV;
solid line, 350 MeV; dash-dotted line, 450 MeV. The branching ratios of the
different decay channels  of the $N^*$ are modified in (c), with the total
width fixed to 350 MeV; dashed line:
$Br(N (\pi\pi)^{T=0}_{S-wave}) = 5 \%$ and $Br(\Delta \pi) = 30 \%$; solid
line: $7.5 \%$ and $25 \%$; dash-dotted line: $10 \%$ and $20 \%$.}
\label{comb}
\end{center}
\end{figure}

Wherever dealing with the Roper resonance, the lack of a precise
determination of its properties is a problem that should be faced. In Fig. 
\ref{comb}, I investigate how the results vary with some of the
uncertainties. As can be noticed, the shape is not affected by these 
changes, but some set of values are preferred.
Fig. \ref{comb} a, shows how the spectrum changes with the
modification of $c^*_1$, $c^*_2$ within the previously accepted values, while
keeping the total width and the partial branching ratios fixed; set II gives a
better agreement with the data. In Fig \ref{comb} b, we show the range of
uncertainties that come from the variation of the total width of the $N^*$
in the limits given by the Particle Data Book \cite{pdg}, that is from 250 MeV
(lower line) to 450 MeV (upper line). Finally, the dependence on the partial
branchings of the $N^*$ to $N (\pi \pi)^{T=0}_{S-wave}$ and $\Delta \pi$,
with the total width fixed to 350 MeV, is studied in Fig. \ref{comb} c.
The dashed line corresponds to a $5 \%$ branching of
$N^*$ to $N (\pi \pi)^{T=0}_{S-wave}$ and a $30 \%$ to $\Delta \pi$, while for
the dash-dotted one, a larger $10 \%$ of $N (\pi \pi)^{T=0}_{S-wave}$ and
a smaller $20 \%$ of $\Delta \pi$; the data prefer the latter choice.
These data alone do not allow to disentangle the different effects, but, in 
combination with other data that will be available in the future (like the 
data on $p p \rightarrow d \pi \pi$ at $T_p = 600 - 775$ MeV from CELSIUS), 
would be an important source of information about the Roper resonance.

\begin{figure}[h!]
\begin{center}
\includegraphics[height=0.5\textwidth]{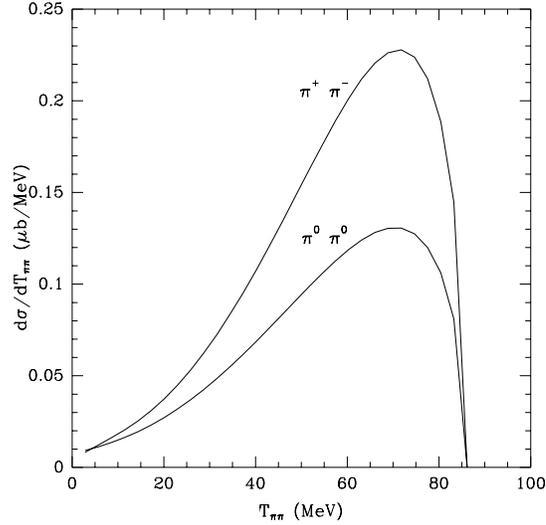}
\caption{Two-pion invariant mass spectrum for $n p \rightarrow d \pi \pi$ at
$p_n = 1460$ MeV for both charged and neutral pions.}
\label{dsdt}
\end{center}
\end{figure}

As mentioned in the Introduction, the present model could provide an
explanation to the deviation from phase space observed in the two pion
invariant mass distribution of the reaction
$p d \rightarrow {^3}He \pi^+ \pi^-$, studied with a proton beam of momentum
$p_p = 1.15$ GeV \cite{momo}. This experimental result has been interpreted
with the hypothesis that the pions are preferably emitted in p-wave \cite{momo1}.
An ansatz based on this idea can explain the data reasonably well but,
unfortunately, the initial tail of the $\rho$ meson produced via
$p d \rightarrow {^3}He \rho$ can hardly produce the required strength. In
Fig. \ref{dsdt}, we show the same observable for the reaction
$n p \rightarrow d \pi \pi$; the shape obtained for the charged pions channel
is very similar to the one observed at COSY. The neutral pion channel, though,
exhibits a shape similar to the one of the charged pions, and only a factor
about 1/2 smaller at the peak position, in apparent contradiction with the 
much smaller rate of $\pi^0$ pairs, compared to the charged ones, obtained
in the MOMO experiment. This is a consequence of the fact that, in
the present model, the pions are almost always produced in s-wave. However, 
we should bear in mind that the interference
pattern can be very different when the nucleons are bound in a ${^3}He$
nucleus instead of a deuteron. Detailed calculations are, therefore, required 
to check if the model can explain the MOMO experiment.

\section{Conclusions}

In summary, a simple model for the $n p \rightarrow d (\pi \pi)^0$ reaction
has been developed, based on a previous model for the free
$N N \rightarrow N N \pi \pi$ and including the most important resonance
contributions. It is shown that the bump in the center of the
deuteron momentum spectra (high $\pi \pi$ masses) observed at a neutron beam
momentum of $p_n = 1.463$ GeV ($T_p = 795$ MeV) can be explained as a
result of the interference of two mechanisms involving the excitation of the
Roper resonance: the dominant and phase-space like
$N^* \rightarrow N (\pi \pi)^{T=0}_{S-wave}$ (Fig. \ref{diag} a) and the
smaller in size $N^* \rightarrow \Delta \pi$ (Fig. \ref{diag} b),
but determinant to obtain the right profile. The mechanism of
double-$\Delta$ (Fig. \ref{diag} c) excitation, considered in an earlier
model for the same reaction, but only including pion exchange, is
significantly reduced by the short range correlations. Other two pion
production reactions like $p p \rightarrow p n \pi \pi$ and
$p p \rightarrow d \pi \pi$, currently studied experimentally at CELSIUS for
energies between 650 and 775 MeV, would be helpful to clarify this issue,
The size of the spectra depends appreciably on
the Roper resonance and can be used, together with other two pion production
reactions, to learn more about this resonance and its decay properties.
Finally, the present model could  be helpful to understand the two-pion
invariant mass distributions observed from $p d \rightarrow {^3}He \pi^+ \pi^-$ 
at COSY.

\section*{Acknowledgements}

The author is indebted to C. Wilkin for drawing his attention to the work of
Hollas and collaborators and its connection with the MOMO experiment. He has
greatly benefited from discussions with E. Hern\'andez, E. Oset and M.
J. Vicente Vacas. He is also grateful to G. Faldt for his hospitality at TSL, 
where part of this work was performed, and acknowledges financial support
from the Generalitat Valenciana. This work has been partially supported by 
DGYCIT contract No.PB 96-0753.

\newpage
\section*{Figures}

\vspace{1cm}

\end{document}